\documentclass[superscriptaddress,twocolumn,showpacs,preprintnumbers,amsmath,amssymb]{revtex4}

\usepackage{graphicx}
\usepackage{subfigure}

\begin{document}

\title{Manipulating magnetism and conductance of an
adatom-molecule junction on metal  surfaces: ab initio study}

\author{Kun Tao}
\affiliation{Max-Planck-Institute of Microstructure Physics,
Weinberg 2, D-06120 Halle, Germany}

\author{V.S. Stepanyuk}
\affiliation{Max-Planck-Institute of Microstructure Physics,
Weinberg 2, D-06120 Halle, Germany}

\author{P. Bruno}
\affiliation{Max-Planck-Institute of Microstructure Physics,
Weinberg 2, D-06120 Halle, Germany} \affiliation{European
Synchrotron Radiation Facility, BP 220, F-38043 Grenoble Cedex,
France }

\author{D.I. Bazhanov }
\affiliation{Faculty of Physics, Moscow State University, 119899
Moscow, Russia}

\author{V.V. Maslyuk}
\affiliation{Martin-Luther-Universit\"{a}t  Halle-Wittenberg,
Institut f\"{u}r Physik, Von-Seckendorff-Platz 1, D-06099 Halle,
Germany}

\author{M. Brandbyge}
\affiliation{Microelectronic Center (MIC), Technical University
of Denmark, DK-2800 Lyngby, Denmark}

\author{I. Mertig}
\affiliation{Martin-Luther-Universit\"{a}t  Halle-Wittenberg,
Institut f\"{u}r Physik, Von-Seckendorff-Platz 1, D-06099 Halle,
Germany}

\begin{abstract}
\label{abstract}

The state of the art ab initio calculations reveal  the effect of
a scanning tunnelling microscopy tip on magnetic properties and
conductance of a benzene-adatom  sandwich on Cu(001). We
concentrate on a benzene-Co system  interacting with a Cr tip. Our
studies give a clear evidence that magnetism and conductance in
molecule-adatom junctions can be tailored by the STM tip.  Varying
the tip-substrate distance the magnetic moment of the Co adatom
can be switched on/off. The interplay between spin-polarized
electron transport  through  the junction and  its magnetic
properties  is demonstrated. A spin-filter effect in the junction
is predicted.

\end{abstract}


\maketitle

The ability to  manipulate the magnetic properties and the
spin-polarized electronic  transport is at the basis of
spintronics. In the past few years, much effort has been devoted
to engineer and study spin-polarized nanostructures at the atomic
level\cite{1,2,3,4,5,6,7,8,9}. The development of the scanning
tunnelling microscope (STM) has opened the field of local
spectroscopy on individual atoms\cite{Crommme1}. Spectroscopic
measurements performed on  magnetic adatoms on metal surfaces have
enabled to resolve  the Kondo effect arising from the interaction
of an adatom with a conduction-electron
continuum\cite{Kondo1,Kondo2,Kondo3}.~The ability to control the
spin-state of an isolated magnetic atom has been recently
demonstrated\cite{SPIN1,SPIN2}. By placing an iron or manganese
atom at a specific location on the copper-nitride thin film,
Hirjibehedin et al., \cite{SPIN1} determined the orientation and
strength  of the anisotropies of individual magnetic adatoms.
Atoms in their experiment can  hold a specific magnetic direction,
which may  allow them to store data. Yayon et al.,\cite{SPIN2}
have used the direct exchange interaction between a single
magnetic atom and a nanoscale magnetic island to fix the spin of
the adatom. The above results are of  great importance for future
atomic-scale technologies and single-atom data storage.
Theoretical and experimental studies have shown that electronic
and magnetic properties of a single adatom on metal surfaces
significantly depend on the tip-surface
distance\cite{Huang,Berndt,Hofer}.

\begin{figure}
\centering
\includegraphics[bb=22 18 650
700,height=2.0in,clip]{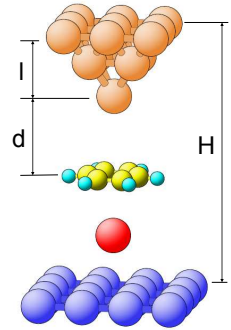} \caption{\label{Dist}(color
online) Setup for the calculations. $H$ is the distance between
the magnetic STM tip and the substrate. $l$ is the length of the
tip, while $d$ is the distance between the benzene molecule and
the tip-apex.}
\end{figure}

Recent experiments of Wahl et al.,\cite{Wahl}~have demonstrated
the ability to tune the spin state of a single magnetic adatom by
the controlled attachment of a molecular ligand. The Kondo
temperature of a Co adatom on Cu(100) was found to significantly
increase with the number of CO molecules attached to the adatom.
Zhao et al.,\cite{Zhao,Crom_zhao} have revealed that the magnetic
state of a cobalt ion trapped within a single phthalocyanine
molecule (CoPc) on Au(111) can be manipulated by the
dehydrogenation of the ligand by voltage pulses from the STM tip.
Recent advances in the field of molecular-spintronics have enabled
the manipulation of spins in molecules with control down to single
spins\cite{Sanvito1,Sanvito2}. Magnetic states of atoms have been
found to significantly change in metal-molecule clusters and
sandwiches\cite{Jena,Zhao,Masl,Xian,Wed,Mokr}. Controlling and
manipulating spins  and conductance of single adatoms and
molecules on surfaces could be of great importance for the
development of quantum nanodevices. Owing to the fascinating
advances in single atom/molecule manipulations with a STM tip, it
is now possible to engineer new nanostructures  in an atom-by-atom
fashion\cite{Eigler,Crommie3,Hla,Gimz,Ho}. The ability of the STM
tip to reversibly modify the internal conformation of a molecule
and  realize molecular switching has been reported\cite{Rieder2}.

In this Letter we point out a way for tailoring the spin and the
transmission  of adatom-molecule junctions  on  metal surfaces
exploiting the tip-molecule  interaction. Performing ab initio
calculations we show that by means of vertical manipulation it is
possible to change  the magnetic moment of an adatom and the
spin-polarized transmission through the junction. We concentrate
on a cobalt adatom sandwiched between Cu(001) and benzene molecule
(Fig.~\ref{Dist}). We demonstrate that magnetism  and
spin-dependent transmission  in such  systems can be controlled by
varying the distance between the molecule and the magnetic tip. A
spin-filter effect in the junction caused by the tip is revealed.

Our calculations are performed  using the density functional
theory(DFT)  and the linear combination of pseudoatomic  orbitals
method implemented in the SIESTA  code\cite{Siesta}. For the exchange
and correlation potential we use both the LDA and the GGA
approximation. Atomic cores are replaced by nonlocal, norm-conserving
scalar-relativistic Troullier-Martins pseudopotentials. An energy
cutoff of 250 Ry is used to define the real-space grid for numerical
calculations involving the electron density. Valence electrons are
described using a double-$\zeta$ plus polarization (DZP) basis set
for Cu and the benzene molecule, and a triple-$\zeta$ plus
polarization (TZP) basis set for the Cr and the Co atoms. The geometries
are optimized until all residual forces on each atom are smaller
than 0.01eV/\AA. The results are confirmed by VASP calculations\cite{VASP,vasp1}.

\begin{table}
\caption{Displacements and magnetic moments of the system for
different tip-substrate separations. $H$ is the tip-substrate
distance, as shown in Fig.~\ref{Dist}. $\Delta l=l-l_{0}$ is the
change of the tip length, where $l$ and $l_{0}$ are the tip length
with and without tip-benzene interaction. $\Delta d=d-d_{0}$ is
the distance difference between Cr tip-apex and the benzene
molecule along $z$ axis after and before full relaxation. The
ideal benzene-Co and Co-substrate distances without tip-benzene
interaction are 1.60\AA~ and 1.48\AA. The last two columns are the
magnetic moment of the Cr tip-apex and Co adatom calculated within
the GGA(LDA) method.} \label{Relaxation}
\begin{center}
\begin{tabular*}{0.48\textwidth}{@{\extracolsep{\fill}} l  c  c  c  c}
\hline
                  H~(\AA)  &  $\Delta l$~(\AA)  &  $\Delta d$~(\AA)  & $M_{Cr}$~($\mu_{B}$)  & $M_{Co}$~($\mu_{B}$)\\
\hline
  $>$10.1   &  0     &  0     &  4.75(4.68)  &    0(0)\\
   9.3      &  0.12  & -0.14  &  4.61(4.61)  &  0.16(0.08)\\
   8.6      & -0.05  &  0.07  &  4.24(4.05)  &  0.64(0.38)\\
   8.1      & -0.25  &  0.28  &  3.67(3.51)  &  0.96(0.58)\\
\hline
\end{tabular*}
\end{center}
\end{table}

In spin-polarized STM experiments, the tip is often made from the
nonmagnetic material coated by thin films of antiferromagnetic
(AFM) or ferromagnetic (FM) materials \cite{1,SPIN2}. In our study, we model the
tip by a pyramid consisting of 13 Cu atoms and one Cr atom for
the tip-apex, as shown in Fig.~\ref{Dist} \cite{TIP}.

In order to investigate the interactions between the STM tip and
the molecule, we first perform \emph{ab initio} calculations to
find an equilibrium position for the benzene molecule adsorbed on
the Cu(001) surface. We use a slab model for the adsorption
system, consisting of 3, 4 and 5 atomic layers with 16 Cu atoms
per layer. Our main results are already well converged for a 3
layers slab. During the geometry optimizations, the structure of
the molecule and the top two layers of the substrate are fully
relaxed but the bottom layer of the substrate is fixed at its bulk
position. Several initial adsorption configurations including
hollow, bridge and atop sites are considered to find the most
stable one. Comparing the total energies of these configurations,
we find that the most stable position of the molecule is over the
hollow site of the surface layer, in agreement with the previous
experimental~\cite{EXP_BENZ} and theoretical~\cite{THEORY_BEZ}
studies.

\begin{figure}
\centering
\includegraphics[width=9cm]{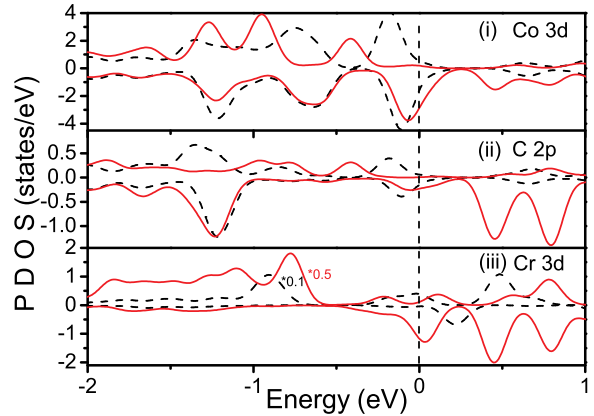}
\caption{\label{PDOS} (Color online) The PDOS of the Co, C and Cr
atoms in C$_{6}$H$_{6}$/Co/Cu(001) junction at two different
tip-substrate
 distances, 9.3\AA~ and 8.1\AA. The black lines are the PDOS at a
 larger tip-substrate distance 9.3\AA~ and the red lines are the
 PDOS at a closer tip-substrate distance 8.1\AA. (i) 3d states of
 the Co atom, (ii) 2p states of the C atoms in the benzene molecule,
(iii) 3d states of the Cr tip-apex. In figure(iii), the values of
th PDOS in the region from -2.0 eV to -0.5 eV are scaled by
factors 0.5 and 0.1, correspondingly.}
\end{figure}

We put a Co atom between the benzene molecule and the Cu(001)
surface, forming a planar sandwich configurations:
C$_{6}$H$_{6}$/Co/Cu(001)\cite{COM1}. Such structures are
experimentally feasible and can be produced using the STM
manipulation\cite{Hla2}.

At different tip-substrate distances, the benzene molecule and the
Co atom show different relaxation behavior, summarized in
Table~\ref{Relaxation}. At a larger tip-substrate separation
9.3\AA, the benzene molecule, Co adatom and the substrate under
the adatom are pulled up, while the Cr tip-apex is pulled down. At
this stage the attractive interactions between the tip and benzene
molecule, and between the Co atom and the substrate are the
driving forces for the observed atomic relaxations. However, at a
closer tip-substrate separation 8.1\AA, the repulsive interactions
between the tip and the benzene molecule, and between the molecule
and the Co atom begin to play an important role. It can be
observed that the benzene molecule, Co adatom and the substrate
are pushed down, while the tip-apex is pushed up.

The changes of the magnetic moment of the Co adatom during the
approach of the STM tip are also summarized in
Table~\ref{Relaxation}. When the tip-substrate distance is larger
than 10.1\AA, the tip  has very weak interaction with the benzene
molecule. Because of the strong hybridization between the C 2p
states and the Co 3d states the magnetic moment of the Co in the
C$_{6}$H$_{6}$/Co/Cu(001) system is quenched to 0$\mu_{B}$. When
the tip-substrate separation decreases from 9.3\AA\ to 8.1\AA, the
magnetic moment of the Co atom increases from 0.16$\mu_{B}$\
(0.08$\mu_{B}$,\ LDA) to 0.96$\mu_{B}$\ (0.58$\mu_{B}$,\ LDA); on
the contrary, the magnetic moment of the Cr tip-apex decreases
from 4.61$\mu_{B}$\ (4.61$\mu_{B}$,\ LDA) to 3.67$\mu_{B}$\
(3.51$\mu_{B}$,\ LDA).

Partial density of states (PDOS) of the Co atom, C atoms in the
benzene molecule and the Cr tip-apex for two different
tip-substrate distances, 9.3\AA~ and 8.1\AA, are plotted in
Fig.~\ref{PDOS}. At a larger tip-substrate distance (9.3\AA), the
hybridization between the 3d states of the Cr atom and the 2p
states of the C atoms is weak. Meanwhile, the hybridization
between the 2p states of the C atoms and the 3d states of the Co
atom is still very strong. The minority part of the Co 3d states
is slightly shifted to the Fermi level which leads to a nonzero
magnetic moment of the Co atom.  However, at a closer
tip-substrate distance (8.1\AA), the 2p states of the C atoms in
the molecule are strongly hybridized with the 3d states of the Cr
and Co atoms. Also, the hybridization between the C 2p states and
Cr 3d states is much stronger than that for the larger distances.
The increased interaction between them pushes the minority 3d
states of the Cr atom to the Fermi level and increases their
population. Therefore, the magnetic moment of the Cr tip-apex
reduces from 4.61$\mu_{B}$ to 3.67$\mu_{B}$. However, one should
note that due to atomic relaxations in the junction (Table 1) the
hybridization between C 2p states and the Co 3d states at a closer
tip-substrate distance is weaker than that at a larger distances.
Therefore the majority part of the Co 3d states moves far away
from the Fermi level~(cf. Fig.2). As a result, the magnetic moment
of the Co atom at a closer tip-substrate distance recovers from
0.16$\mu_{B}$ to 0.96$\mu_{B}$.

To gain detailed insight into the effect of the tip on the transport
properties of the junction, we
have performed transport calculations using the
TranSiesta~\cite{TRANS} code, where the non-equilibrium Green
function method is implemented. The results have been calculated
with 64 energy points. The bottom of the
valence band was at -6~Ry due to the presence of the pseudovalent 3$p$
states. Details of transport calculations can be found in
\cite{TRANS}.

\begin{figure}[b]
\centering
\includegraphics[width=9cm]{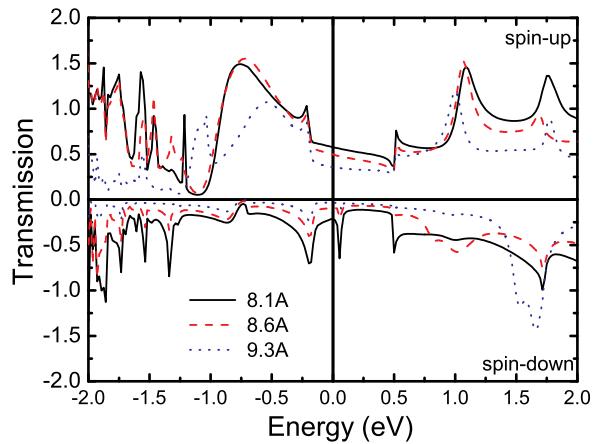}
\caption{\label{Transm}(color online) The transmission spectra for
three different positions of the STM tip. Positive values related
for the spin-up channel and negative correspond to the spin-down.}
\end{figure}

Fig.~\ref{Transm} shows spin-resolved transmission probabilities
through the Co-benzene molecule for three different positions of
the STM tip. One can see  that with decreasing the distance
between the tip and the substrate the transmission  at zero bias
increases. Increasing the distance between the tip leads to
quenching of the transmission for the spin-down channel. These results
clearly show that such junctions can be used as a well controlled
spin-filter.

In conclusion, our findings have demonstrated the ability  to
tune the  spin and the transport properties of a metal-molecule junction
by the STM tip. We have shown that the electronic, magnetic and
electrical properties of the Co-benzene junction on a Cu(001) surface can be
manipulated by changing the tip-substrate distance.
Spin-selectivity in transmission can be achieved by an appropriate
choice of the position of the STM tip. The physics behind all
effects found in this work is related to atomic relaxations in the
junction caused by the interaction with the tip. Therefore, we
expect that similar effects can be detectable with current
technology for different metal-molecule junctions.


\begin{thebibliography}{0}
\bibitem{VASP}
To confirm spin-polarized SIESTA calculations, we have applied
VASP~\cite{vasp1} in an all-electron projector augmented-wave
(PAW) representation \cite{PAW} of electron-ion interaction
potentials. This approach shares the computational efficiency of
the pseudopotential technique, but is an all-electron scheme
providing very realistic valence-electron distributions important
for studing the magnetic properties of the system. We have found
that PAW method reproduced magnetic calcultions carried out with SIESTA,
but with less local magnetic moments with regard to the choice of
the atomic Wigner-Seitz radius (with difference less than 0.2
$\mu_B$ per Co and Cr atoms). In VASP calculations the electronic
states are expanded in terms of plane waves with the number
restricted by maximal kinetic energy cut-off of 400 eV employed in
present calculations.
\bibitem{TIP}   Since Cr tip is an AFM tip, we also perform calculations for  the
tip with different sizes of antiferromagnetic Cr clusters placed  at its apex.
Results obtained in both approaximations are nearly the same.
\bibitem{COM1} The possible inclination of the benzene molecule above Co atom towards
the Cu(001) surface plane was considered as well, but it was found to
be less favourable than the planar configuration (with gain of
energy of about 0.1 eV).
\end{thebibliography}
\end{document}